\pdfoutput=1

\documentclass[]{raa}

\usepackage{mathptmx}
\usepackage{ae,aecompl}

\usepackage{graphicx,times}	
\usepackage{amsmath}	
\usepackage{amssymb}	
\usepackage{pifont}
\usepackage{txfonts}
\usepackage{url}
\usepackage{subfigure}
\usepackage{longtable}
\usepackage{lscape}
\usepackage{multirow}
\usepackage{color}
\usepackage{textcomp}
\usepackage{natbib}
\usepackage{epstopdf}
\usepackage{pdfpages}
\usepackage[colorlinks,linkcolor=red,anchorcolor=green,citecolor=blue]{hyperref}
\usepackage{soul}
\usepackage{appendix}

\newcommand{\cg}{\color{green}}
\newcommand{\bg}{\bf\cg}

\begin{document}

\title{ The Multiple Images of the Plasma Lensing FRB}


   \volnopage{Vol.0 (2021) No.0, 000--000}      
   \setcounter{page}{1}          
   \author{Yu-Bin Wang
      \inst{1,2}
   \and Zhi-Gang Wen
      \inst{1,3,4}
   \and Rai Yuen
      \inst{1,3}
   \and Na Wang
      \inst{1,3,4}
   \and Jian-Ping Yuan
      \inst{1,3,4}
   \and Xia Zhou
      \inst{1,3,4}
   }

 \institute{Xinjiang Astronomical Observatory, Chinese Academy of Sciences, Urumqi, XinJiang 830011, China \\
        \and
            University of Chinese Academy of Sciences, 19A Yuquan Road, Beijing 100049, China \\
        \and
            Key Laboratory of Radio Astronomy, Chinese Academy of Sciences, Nanjing 210008, China\\
        \and
            Xinjiang Key Laboratory of Radio Astrophysics, 150 Science1-Street, Urumqi, Xinjiang, 830011, People's Republic of China\\
\vs\no
   {\small Received 20xx month day; accepted 2022 April 22th}}
\email{{\it wenzhigang@xao.ac.cn; ryuen@xao.ac.cn; na.wang@xao.ac.cn}}

\abstract{We investigate the formation of multiple images as the radio signals from fast radio bursts (FRBs) pass through the plane of a plasma clump.
The exponential model for the plasma clump is adopted to analyze the properties of the multiple images.
By comparing with the classical dispersion relations, we find that one image has exhibited specific inverse properties to others, such as their delay times at high frequency is higher than that at low frequency, owing to the lensing effects of the plasma clump.
We demonstrate that these inverse effects should be observable in some repeating FRBs.
Our results predict deviation in the estimated DM across multiple images, consistent with the observations of FRB 121102 and FRB 180916.J0158+65.
If other plasma lenses have effects similar to an exponential lens, we find that they should also give rise to the similar dispersion relation in the multiple images.
For some repeating FRBs, analysis of the differences in time delay and in DM between multiple images at different frequencies can serve as a method to reveal the plasma distribution.
\keywords{galaxies: distances and redshifts $-$ galaxies: interstellar medium $-$ gravitational lensing: strong $-$ ISM: structure $-$ scattering}}

   \authorrunning{Yu-Bin Wang, et al.}                     
   \titlerunning{The Effects of Plasma Lens}       

   \maketitle


\section{Introduction}\label{Sect.1}

Fast radio bursts (FRBs) are bright extragalactic transient radio pulses, in the order of Jansky, with durations of a few milliseconds. The first FRB, also known as the Lorimer burst (\citealt{Lorimer2007}), was discovered in 2007 in the Parkes radio telescope archival data.
Since then, more than 600 FRBs have been detected by many telescopes around the world (\citealt{Petroff2016,Luo2020,Amiri2021}).
Among them, 27 FRBs have been reported with multiple bursts, and eleven have been given the exact locations (\citealt{Connor2020,Macquart2020}).
By comparing the free electron column density derived along the line of sight to the FRBs with that in the Milky Way, anomalously high dispersion measures (DMs) were obtained for FRBs.
This indicates that they are extragalactic sources (or cosmological origin) rather than Galactic origin (\citealt{Thornton2013}). FRBs can be broadly categorized as repeating and non-repeating. The origin of repeating FRBs, or repeaters for short, may be distinctly different from that of the non-repeating FRBs (\citealt{Andersen2019}), or non-repeaters. For instance, the emission mechanism in the former has been suggested in relation to the luminous coherent emission processes around magnetars (\citealt{Kumar2017,Andersen2019,Andersen2020,Li2021a}).
An example is that found in SGR 1935+2154 (FRB 20200428) in the Milky Way, which possesses several features similar to repeaters. On the contrary, catastrophic events such as collapse of compact objects and supernovae have been associated with the cause of the non-repeaters (\citealt{Platts2019}). Many theories have been proposed (\citealt{Platts2019,Zhang2020,Xiao2021}) but the origin of FRBs remains one of the popular investigations in science.

Radio signals from large cosmological distances are dispersed when propagating through cold plasma.
In the classical form, the delay time, $t_{\rm d}$, is related to the DM and the frequency of the signal, $\nu$, given by $t_{\rm d} \propto \nu^{-2} \rm DM$, where ${\rm DM} = \int{n_{\rm e} \, dl}$ represents the free electron column density along the line of sight.
In general, the electron density is dependent on the propagation path of the radio signal.
This gives rise to the plasma lensing effects, such as diverging refraction (\citealt{Clegg1998}), resulting in multiple images and delays in the received signals.
Similarly, signals from some repeaters can also suffer from the effects of plasma lens resulting in possible multiple images (\citealt{Cordes2017,Er2018}), with the delay times showing unusual time$-$frequency relation after de-dispersion (\citealt{Tuntsov2021}).
Such delay times in bursts have been reported in the observations of some repeaters (\citealt{Gajjar2018,Shannon2018,Andersen2019,Amiri2019,Fonseca2020}).
They exhibit as downward drift in the observing frequency in a sequence of bursts known as ``sad trombone''.
The radius-to-frequency mapping, which suggests that radiation observed at different frequencies are coming from different heights, can only explain delays of several milliseconds between bursts (\citealt{Wang2019,Lyutikov2020}).
However, some repeaters emit independent pulses in time interval of about tens of milliseconds (\citealt{Chawla2020,Platts2021}).
Another type of delay times observed from bursts of some repeaters' bursts exhibits as upward drift in frequency or ``happy trombone'' (\citealt{Rajabi2020}).
In addition, the measured DMs are low at low frequencies compared to that at high frequencies.
For example, the DM difference at frequency between $0.9-1.6 \, \rm GHz$ is approximately 1--2\,$\rm pc \, cm^{-3}$ for FRB 121102 (\citealt{Platts2021}), and the difference in DM is $\gtrsim 0.5 \, \rm pc \, cm^{-3}$ for FRB 180916.J0158+65 at frequency between $0.4-0.8 \, \rm GHz$ (\citealt{Chamma2020}).
This is different from that suggested by the radius-to-frequency mapping model (\citealt{Wang2019,Lyutikov2020}). It is also incompatible with a gravitational lens, which demonstrates $\lesssim 10 \, \rm ms$ delay between the lensed images from the burst (\citealt{Munoz2016}).

The effects of a plasma lens are determined by parameters such as the characteristic scale and the plasma density along the line of sight as well as the frequency of bursts.
These parameters are dominant in the plasma lens that forms multiple imaged bursts.
When the emission from repeating FRBs passes through a plasma lens of large structure, the delay times due to the geometric effect dominates, which can account for the formation of the ``sad trombone" (\citealt{Er2020}).
For high magnification ($\mu > 5$), multiple images at the same frequency will have different arrival times ranging from less than a few microseconds to tens of milliseconds (\citealt{Cordes2017}).
However, the spectral pattern will appear to be very narrow-band, which is different from that observed in the FRBs (\citealt{Gajjar2018,Pastor-Marazuela2020}).
The results given by \cite{Cordes2017} are also insufficient to explain several inverse properties observed in some multiple images that vary across the frequency bandwidth compared with other images.
For example, their delay times may display as ``happy trombone'' in opposite to the behavior mentioned by \citet{Er2020}.
From the observations (\citealt{Chatterjee2017,Tendulkar2017}), FRB 121102 has been associated with a persistent radio and optical source possessing a projected size of $\lesssim 0.7\, \rm pc$. Multiple images observed from FRB 180916.J0158+65 may also originate from the effect of a plasma lens (\citealt{Amiri2020}).
The FRB is located behind a star-forming clump with a projected size of roughly $1.5\, \rm kpc$, and the source environment occupies the whole clump with the structure spanning between $30\sim 60\,\rm pc$ (\citealt{Marcote2020,Tendulkar2021}).
The latent plasma lens may be hidden behind the clump.
In addition, the circular polarisation of up to $75\%$ and the source environment in FRB 20201124A suggest that the radiation from the repeater may pass through a foreground object before reaching the observer (\citealt{Xu2021}).
Many repeaters that discovered at frequencies between $400-800 \rm \, MHz$ also show time delay characteristics in the images similar to that from FRB 180916.J0158+65 (\citealt{Amiri2019,Fonseca2020}).
This paper will investigate the possibility of the formation of multiple images due to a plasma lens.
The frequency-dependent delay time from the multiple images would cause bias in the observed dispersion relation of the FRB, and we will discuss the relationship between the delay times in multiple images and the dispersion relation.

The paper is organized as follows.
In Section 2 we outline the theory and the equations for plasma lens.
In Section 3 we will discuss the possible effects from a plasma lens of exponential form.
Discussion and a summary of the paper is given in Section 4.
In this paper, the parameters for the standard $\Lambda$CDM cosmology are adopted as $\Omega_{\Lambda} = 0.6791$ and $\Omega_{\rm m} = 0.3209$ based on the Planck data, and the Hubble constant is taken as $H_0 = 100 \,h \rm \, km \, s^{-1} \, Mpc^{-1}$ with $h = 0.6686$ (\citealt{Adam2016}).

\section{The Basic Model of Plasma Lens}\label{Sect.2}

\begin{figure}
\centering
  \includegraphics[width=0.45\textwidth, angle=0]{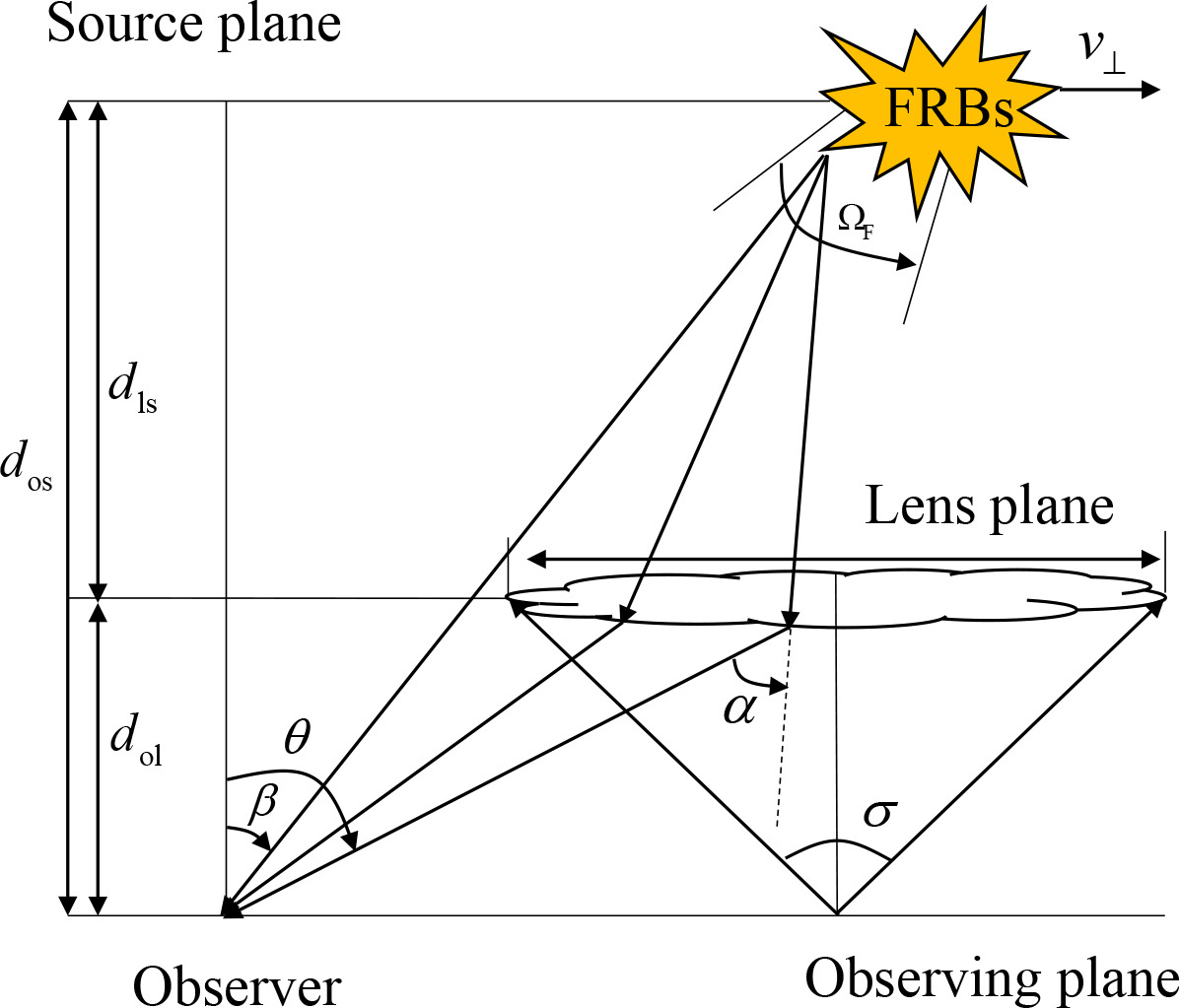}\\
  \caption{Diagram showing refracted light rays by a plasma lens. The $\sigma$ represents the effective angular structure of the lens as defined by \citet{Vedantham2017}.}\label{FRB}
\end{figure}

We assume a thin and axially symmetric lens in Cartesian angular coordinate system.
The geometric effect for light rays passing through the thin plasma lens can be expressed in the equation similar to that due to a gravitational lens (\citealt{Schneider1992}).
Fig. \ref{FRB} shows the geometry for deflected light rays from FRBs illustrating the additional geometric and dispersive delays as compared to non-refractive bursts.
In this model, the distribution of the deflected rays on the image plane is described by the gradient of the deflection potential given by (\citealt{Schneider1992})
\begin{eqnarray}
\beta = \theta  -  \alpha  =  \theta  -  \nabla_{\theta} \psi(\theta). \label{beta}
\end{eqnarray}
Here, $\beta$ and $\theta$ are the angular positions of the source and the image, respectively, and $\alpha$ is the deflection angle of the light ray due to the plasma lens. The deflection potential of the lens is signified by $\psi(\theta)$, and $\nabla_{\theta}$ represents the gradient with respect to the angular position on the image plane.

The deflecting structure of the plasma lens is described as a refractive medium with spatially varying refractive index. The deflection potential is due to perturbation in the effective refractive index (\citealt{Wagner2020}), which results in greater phase velocity through the lens than the speed of light, $c$, in vacuum.
The $\psi(\theta)$ is related to the dispersive potential, $\tilde{\psi}(\theta)$, by Fermat's principle (\citealt{Wagner2020}), which can be summarized as (\citealt{Fiedler1987,Cordes2017,Vedantham2017,Wagner2020})
\begin{eqnarray}
\nonumber \psi(\theta)  &=&    \frac{d_{\rm ls}}{d_{\rm os} d_{\rm ol}} \tilde{\psi}(\theta)\\ \label{psi}
                        &=&   \frac{1}{(1 + z_{\rm d})} \frac{d_{\rm ls}}{d_{\rm os} d_{\rm ol}} \frac{r_{\rm e} c^2}{2\pi \nu^2} N(\theta).
\end{eqnarray}
Here, $ d_{\rm ls}$ is the distance between the lens and the source, $d_{\rm os}$ is the distance from the observer to the source, and $d_{\rm ol}$ represents the distance from the observer to the lens.
The classical radius of an electron is given by $r_{\rm e}$, $z_{\rm d}$ is the redshift at the lens plane, and $N(\theta)$ is the projected electron density along the line of sight. We assume $N(\theta)\approx \rm DM(\theta)$, where $\rm DM(\theta)$ specifies the DM of the lens at $\theta$.
For large distances and approximating light rays reaching the lens in parallel, the beaming solid angle of FRB ($\Omega_{\rm F}$), as given by \citet{Kumar2017}, is much greater than the effective solid angle of the plasma lens ($\sigma$).

By comparing with the non-lensing case, the total delay time, $t_{\rm tot}$, is the sum of the dispersive and geometric delays. The geometric delay, $t_{\rm g}$, is due to the increased path of propagation along the trajectory from the source position to the observer, and the dispersive delay, $t_\psi$, is owing to the increased DM in the path of propagation.
They are given by (\citealt{Cordes2017,Wagner2020})
\begin{equation}
t_{\rm g}    = \frac{1}{2} \frac{(1 + z_{\rm d})}{c} \frac{d_{\rm os} d_{\rm ol}}{d_{\rm ls}} \alpha^2(\theta,\nu), \label{geometric_delay}
\end{equation}
and
\begin{equation}
t_\psi =  \frac{(1 + z_{\rm d})}{c} \frac{d_{\rm os} d_{\rm ol}}{d_{\rm ls}}\psi(\theta,\nu). \label{dispersive_delay}
\end{equation}
Coupled with Equations (\ref{beta}) and (\ref{psi}), the geometric delay has a relationship signified by $\alpha^2(\theta,\nu) \propto \rm DM(\theta)^2 \, \nu^{-4}$, and the dispersive delay is given by $\psi(\theta,\nu) \propto \rm DM(\theta) \, \nu^{-2}$.

The plasma lens may be located in the Milky Way, in the host galaxy of the FRB or in faint intervening galaxies in intergalactic space (\citealt{Vedantham2017,Er2020}). This leads to time delay and perturbation in the DM in the observed bursts from repeating FRBs as revealed by the multiple images that are caused by plasma lenses.
As small perturbations in DM have been reported in FRBs 180916.J0158+65 and 121102 (\citealt{Amiri2020,Li2021b}), in the following sections, we will discuss multiple images as due to the effects of a plasma lens at different distances and different effective structures.
Based on the suggested possible source distance of around $d_{\rm os} \approx 1 \, \rm Gpc$ (\citealt{Petroff2016,Amiri2021}), we compare the differences in the properties of the multiple images assuming that the plasma lens is located in (i) the host galaxy of the FRB ($d_{\rm ls} \approx 1 \, \rm kpc$), (ii) the faint intervening galaxy ($z_{\rm d} \approx 0.0219367 \sim d_{\rm ol} \approx 100\, \rm Mpc$), and (iii) the Milky Way ($d_{\rm ol} \approx \rm 1 \, kpc$).
Although the axially symmetric electron distribution within a plasma lens has been widely described in the exponential and the power-law models (\citealt{Clegg1998,Cordes2017,Vedantham2017,Er2018,Er2020}), there is still lack of a detailed empirical or analytical expression for the density structure of the plasma.
The similar multiple images can be predicted from the two models.
However, the power-law model requires removal of a singularity in the electron density at the center of the lens, and the addition of a finite core with angular core radius ($\theta_{\rm C}$) to the angular radius $\theta \sim (\theta^2 + \theta^2_{\rm C})^{1/2}$ (\citealt{Er2018}).
Considering the parameter $\theta_{\rm C}$ in the power-law model being artificial and also influential for determining whether multiple images will be produced by the lens (\citealt{Er2018}), we will adopt the exponential lens as it is sufficient to interpret the observations as shown in the next section.

\section{The Multiple Images}\label{Sect.3}

\subsection{Multiple images due to an exponential lens}\label{Sect.3.1}

A special case of the exponential model involves the axisymmetric Gaussian lens ($\rm h=2$) (\citealt{Clegg1998}), which was introduced to describe the U-shaped trough observed in some extragalactic sources.
The other exponential forms ($\rm h=1$ and $\rm h=3$) have been developed by \citet{Er2018}.
In this model, a single lens is considered along the line of sight in order to study the distinct physics graph. The exponential form for DM in the plane of the lens is given by (\citealt{Clegg1998,Vedantham2017,Er2018,Rogers2019})
\begin{eqnarray}
\rm DM (\theta)  = \rm DM_0 \, \exp\bigg( -\frac{\theta^h}{h \sigma^h}\bigg), \label{DM}
\end{eqnarray}
where $\rm DM_0$ represents the maximum electron column density of the lens.
Using Equations (\ref{psi}) and (\ref{DM}), the deflection potential can be rewritten as
\begin{eqnarray}
\psi(\theta) = \theta_0^2 \rm \exp\bigg( -\frac{\theta^h}{h \sigma^h}\bigg),    \label{psi2}
\end{eqnarray}
where $\theta_0$ is the characteristic angular scale which has the form given by
\begin{eqnarray}
\theta_0 (\sigma,\nu,\rm DM_0)  &=&  \bigg[\frac{1}{(z_{\rm d} + 1)} \frac{d_{\rm ls}}{d_{\rm os} d_{\rm ol}} \frac{r_{\rm e} c^2}{2\pi \nu^2}  \rm DM_0 \bigg]^{1/2}. \label{theta_0}
\end{eqnarray}
To simplify the calculations, the exponential forms defined by $\rm h=1$, $\rm h=2$ and $\rm h=3$ as referred by \citet{Er2018} are discussed in the next paragraph.

\begin{figure*}
\centering
  \includegraphics[width=0.99\textwidth, angle=0]{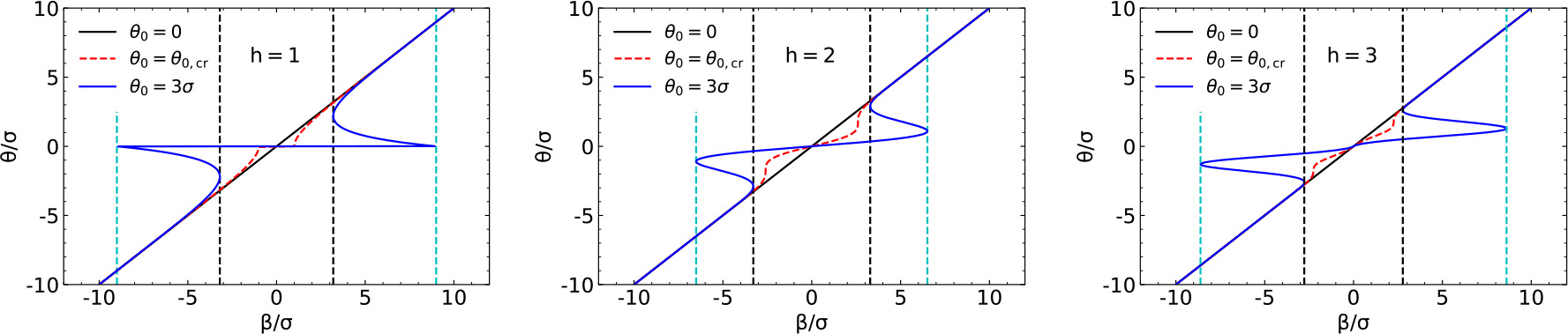}\\
  \caption{Plots showing the relationships between image position ($\theta$) and source position ($\beta$) for three different values of $\theta_0$. In each plot, the center of the lens is at $\theta = \beta = 0$, and the dashed black and dashed cyan lines represent, respectively, the inner and outer boundaries between which multiple images occur.}\label{beta_theta_sigma_x}
\end{figure*}

\begin{figure*}
\centering
  \includegraphics[width=0.70\textwidth, angle=0]{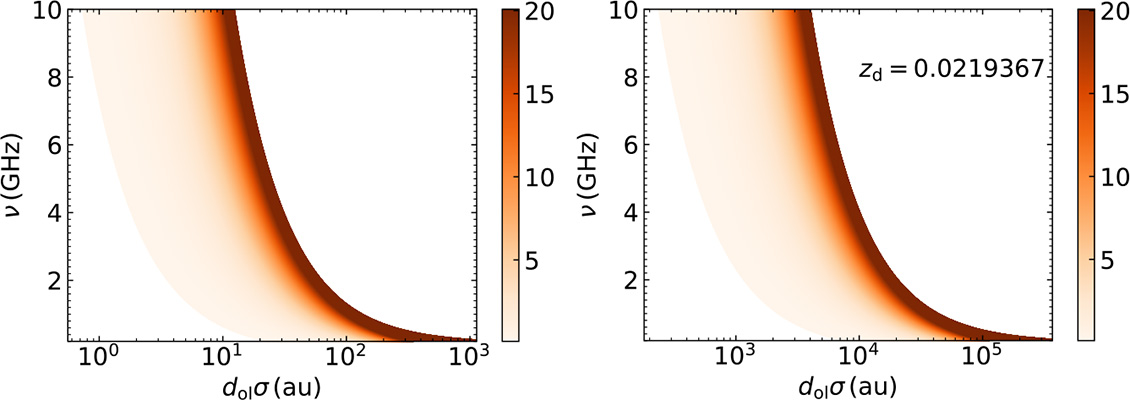}\\
  \caption{Plots showing the relations of $\rm DM_0$, $\sigma$ and frequency at the critical value ($\theta_{0,\rm cr} = \frac{\sqrt{2}}{2}\exp(3/4)\sigma$). The frequency range is from $0.1\,\rm GHz$ to $\rm 10 \, GHz$. The orange region represents different $\rm DM_0$ that scales from $0.1 \, \rm pc \, cm^{-3}$ to $20 \, \rm pc \, cm^{-3}$. The plasma lens on the left is based on $d_{\rm ls} = \rm 1\, kpc$ and $d_{\rm ol} \approx d_{\rm os} = \rm 1\, Gpc$ (in the host galaxy), and the plasma lens on the right is located in the faint intervening galaxy with $z_{\rm d} = 0.0219367$.}\label{sigma_nu_DM}
\end{figure*}

Formation of multiple images requires the partial derivative of Equation (\ref{beta}), with respect to $\theta$, to satisfy $1/\partial_{\theta}(\beta) < 0$.
This means that the minimum characteristic angular scale for $\rm h=1$, $\rm h=2$ and $\rm h=3$ each corresponds to the critical value of $\theta_0$ given by $\theta_{0,\rm cr} = \sigma$, $\theta_{0,\rm cr} = \frac{\sqrt{2}}{2}\exp(3/4)\sigma$ and $\theta_{0,\rm cr} = [(\sqrt{7} +1)^{-1/2}(\sqrt{7} + 3)^{-1/6} \rm \exp(\frac{3 + \sqrt{7}}{6})] \sigma$, respectively, and $\theta_{0,\rm cr}^{\rm h=2} > \theta_{0,\rm cr}^{\rm h=3} \approx \theta_{0, \rm cr}^{\rm h=1}$.
The Young diagrams as defined by Equations (\ref{beta}) and (\ref{psi2}) are given in Fig. \ref{beta_theta_sigma_x}.
In each of the three plots, the unlensed case ($\theta_0 =0$) is signified by the black solid line, and the red dashed curve corresponds to $\theta_0 = \theta_{0,\rm cr}$.
In addition, the case for emerging multiple images is represented by the blue solid curve ($\theta_0 >\theta_{0, \rm cr}$).
The curve has two critical turning points in each of the positive and negative $\beta$ ranges, which are marked by the vertical cyan and black dashed lines, illustrating the dual-caustic structure. We refer to the areas enclosed by the two cyan and the two black dashed lines as windows of multiple images and the corresponding dashed lines indicate the outer and inner boundaries, respectively.
For $\theta_{0} > \theta_{0,\rm cr}$, a source locates between the two boundaries results in two (for $\rm h=1$) or three image positions implying that two or three images are detectable, whereas only one image is obtained from the plasma lens for $\theta_{0} \leq \theta_{0,\rm cr}$.
It is apparent from Fig. \ref{beta_theta_sigma_x} that the outermost image from the lens center has $\theta \approx \beta$, and the positions of other images deviate from the source position.
The figure also shows that image deflection due to the lens with $\rm h=1$ or $\rm h=3$ is stronger than Gaussian lens for identical $\theta_0$, and the two lenses show similarities to the Gaussian lens. This suggests that only Gaussian lens is required to account for the multiple images, and we will consider only the case of h=2 for the rest of the paper.
From Equation (\ref{theta_0}), the $\theta_{0}$ is related to the observing frequency and DM, such that $\theta_{0} \propto\nu^{-1}$ and $\theta_{0} \propto{\rm DM_0^{1/2}}$.
Fig. \ref{sigma_nu_DM} demonstrates the relationships of these parameters at the critical value for $\rm h=2$ ($\theta_{0,\rm cr} = \frac{\sqrt{2}}{2} \exp(3/4)\sigma$). It indicates that the multiple images are constrained by $\sigma$, $\nu$ and $\rm DM_0$.
For specific values of $\rm DM_{0}$ and $\sigma$, and assuming that the multiple images begin at $1 \, \rm GHz$, i.e., $\theta_{0,\rm GHz} = \theta_{0,\rm cr}$, multiple images will also appear at lower frequency.

The diverged images due to the plasma lens yield either a burst of amplification or attenuation, which is inversely related to the determinant of the Jacobian matrix defined by $\mu^{-1} = \rm det(\emph{\textbf{A}})$, where $\emph{\textbf{A}} = \partial \beta/\partial \theta$.
The inverse magnification from the potentials of an exponential lens is given by \citep{Er2018}
\begin{eqnarray}
\nonumber \mu^{-1} &=& \rm 1 + h \theta_{0}^2 \frac{\theta^{h-2}}{\sigma^{h}} \bigg(1 - \frac{\theta^h}{h\sigma^h}\bigg) e^{-\frac{\theta^h}{h\sigma^h}}  \\
         & & + \rm \theta^4_0 \frac{\theta^{2(h-2)}}{\sigma^{2h}} \bigg(h - 1 - \frac{\theta^h}{\sigma^h} \bigg) e^{-2\frac{\theta^h}{h\sigma^h}}. \label{magnification}
\end{eqnarray}
From Equation (\ref{magnification}), the magnification for each image is subjected to $\theta_{0}$ and the image position.
From Fig. \ref{beta_theta_sigma_x}, the positions for the multiple images change as the source position varies.
We refer to the variation in the positions of the multiple images across the plots from large to small as the first, second and third images, respectively.
The first image with the largest image position leads to $\exp[-(\theta/\sigma)^{\rm h}] \rightarrow 0$ in Equation (\ref{magnification}), and the magnification is estimated to be $\mu \simeq 1$.
However, the second and third images at some source positions can have much lower magnifications suggesting that only one image is detectable.
Based on the observed intensity density ratio of the multiple images (\citealt{Amiri2020,Platts2021}), the minimum magnification of all images is set to $\mu = 0.1$ hereafter.

It should be mentioned that not only does the images caused by a plasma lens come from different propagation paths, they also suffer from different DMs as $\theta$ is different. These lead to the different delay times in different images.
From the Young diagrams and Equation (\ref{psi2}), the deflection potential of the first image satisfies $\nabla_{\theta}\psi(\theta) \approx \psi(\theta) \approx 0$, meaning that the first image should have relatively shorter delay time and lower DM than that of the other two images.
It also indicates that the differences in the delay time and DM will be present in between the first and the other two images.
The image positions of the first and third images increase monotonically as the source position increases, but decreases for the second image.
This suggests that specific properties in the second image should be in opposite to that in the first and third images.
These properties will be examined in the next subsection.

\subsection{The delay times and DMs for each image}\label{Sect.3.2}

For a plasma lens that forms multiple images, the lensing parameters are not only constrained by the critical value shown in Fig. \ref{sigma_nu_DM} but their values are also required to take the observations into consideration.
Pulsar observations give the size for a diverging plasma lens ranging from one au to tens of au in the Milky Way, and possibly larger (\citealt{Graham Smith2011,Kerr2018}). The lens, in the environment of the repeaters, may have similar structure to that in the Milky Way.
However, the size of the plasma lens in the intervening galaxy is likely to be much greater than that in the Milky Way or in the host galaxy (\citealt{Vedantham2017,Er2020}), otherwise the effects of the lens will be insignificant for $\sigma \rightarrow 0$ (\citealt{Wagner2020}).
Here, we assume a small scale Gaussian lens with either $d_{\rm ol} \sigma = 30 \, \rm au$ or $d_{\rm ol} \sigma = 50 \, \rm au$ in both the host galaxy and the Milky Way, and either $ d_{\rm ol} \sigma = 10^4 \, \rm au$ or $d_{\rm ol} \sigma = 2 \times 10^4 \, \rm au$ in the intervening galaxy.
In addition, most repeaters were discovered between around $\rm 400\, MHz$ and $\rm 800\, MHz$ by CHIME or even lower at $100\, \rm MHz$. Their extragalactic DMs are in the range between 60 and $\sim 3000 \rm \, pc \, cm^{-3}$ (\citealt{Amiri2021}), with the estimated DM for FRB 180916.J0158+65 being $149.2 \, \rm pc \, cm^{-3}$ (NE2001) or $19.2 \, \rm pc \, cm^{-3}$ (YMW16) (\citealt{Andersen2019}).
From the study of our Galactic halo (\citealt{ProchaskaZheng2019}), the contribution of DM from intervening galaxy is expected in the range of $\approx 50-80 \, \rm pc \, cm^{-3}$.
Based on the discussion above, we assume $\rm DM_0 = 10 \, pc \, cm^{-3}$. In addition, the observed characteristics in the radio signals are likely due to a collective effect of multiple plasma lenses. In this paper, we consider only the case of a single lens.

\begin{figure*}
\centering
  \includegraphics[width=0.99\textwidth, angle=0]{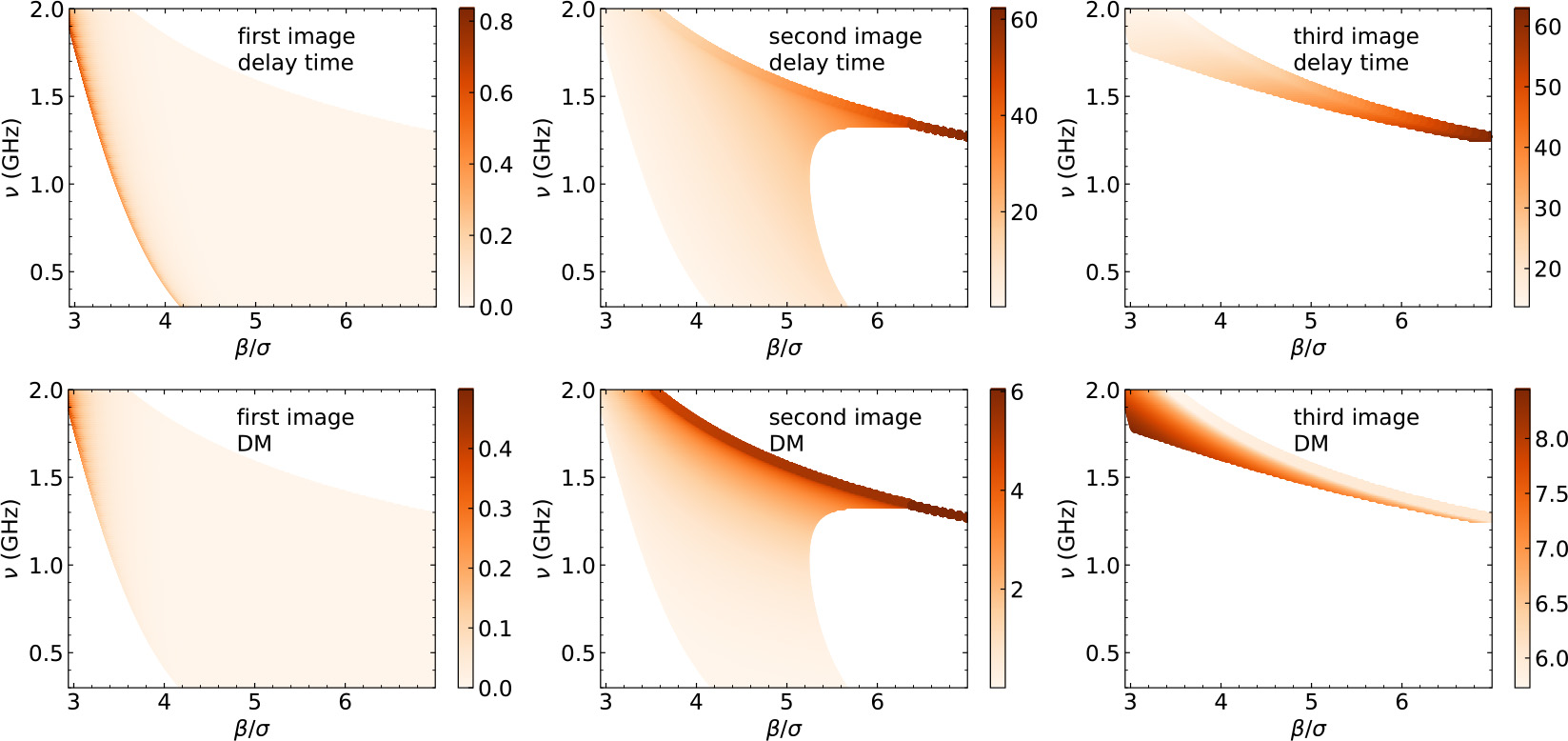}\\
  \caption{Delay times (upper panel) and DMs (lower panel) for the three images caused by a Gaussian lens. The orange regions represent the delay time (in milliseconds) and DMs (in $\rm pc \, cm^{-3}$) for frequencies between $0.3-2.0 \, \rm GHz$. The plasma lens is assumed at $d_{\rm ls} = \rm 1\, kpc$, $d_{\rm ol} \approx d_{\rm os} = \rm 1\, Gpc$ (in the host galaxy) with $d_{\rm ol} \sigma = 30 \, \rm au$ and the magnification $\mu \geq 0.1$.}\label{delay_x_solution}
\end{figure*}

\begin{figure*}
\centering
  \includegraphics[width=0.99\textwidth, angle=0]{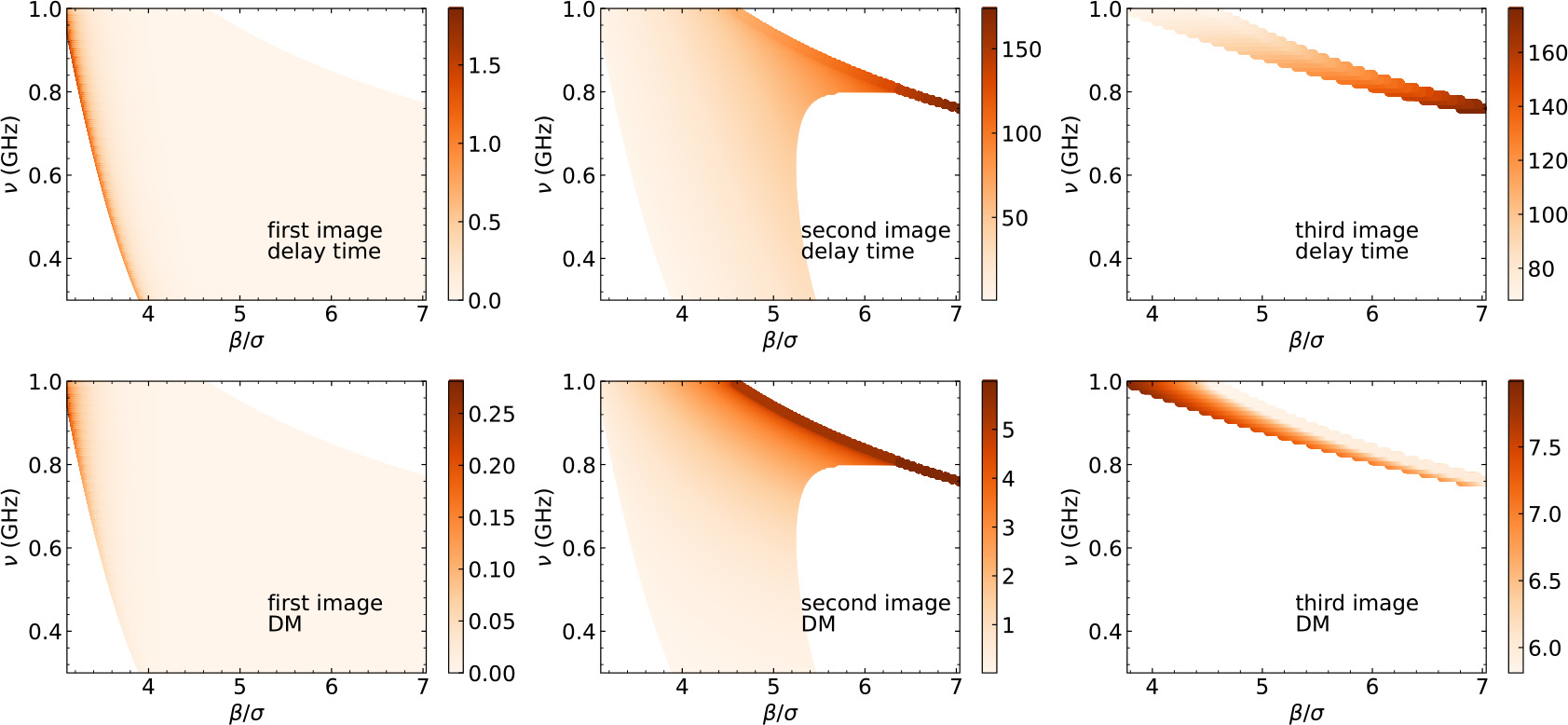}\\
  \caption{Similar to Fig. \ref{delay_x_solution}, but with the structure scale of the plasma lens assumed at $d_{\rm ol} \sigma = 50 \, \rm au$. The frequency is given in the range of $0.3-1.0 \, \rm GHz$.}\label{delay_x1_solution}
\end{figure*}

Coupled with Equations (\ref{beta})$-$(\ref{magnification}), the variations in the delay time and in the DM for each image due to the Gaussian lens in the host galaxy are shown in Figs. \ref{delay_x_solution} and \ref{delay_x1_solution}.
From the two figures, the observable range of frequency for each image from a specific source position is limited by the boundaries at high and low frequencies, which is referred to as the frequency window.
It can be seen that all frequency windows for multiple images exhibit downward drift to lower frequencies as the source position increases.
The frequency at which the first image is observable is dominated by the dual-boundary, and the drift rate at the higher boundary is greater than that at the lower boundary.
The frequency windows for the second and third images are each limited by the given magnification.
Firstly, they also exhibit changes with increasing source position, similar to that seen in the first image.
However, the second image bifurcates into two bandwidths at high and low frequencies as $\beta$ increases, whereas the bandwidth of the third image becomes narrower as frequency decreases.
The delay time and DM are also dependent on the source position and observing frequency.
Delays in the second and third images are much longer than $\sim 1 \rm \, ms$, which is different from the first image.
For larger source position, there exists certain frequency range where both the second and third images possess longer delay times.
The first and third images at specific source positions have much longer delay times at lower frequencies, whereas the second image shows increasingly shorter delay time as the frequency decreases.
For a given source position, the DM in each of the first and third images increases as the frequency decreases, but it decreases in the second image.
The third image has higher DM than that in the other images, with the maximal DM in the first image being lower than $0.5\, \rm pc \,cm^{-3}$.
Figs. \ref{delay_x_solution} and \ref{delay_x1_solution} also show that the window of multiple images is downward drifting to lower frequencies as the effective angular structure increases.
The second image at the same source position and same observing frequency has higher delay time and DM as it passes through a plasma lens with greater effective angular structures.
Similar geometric effects due to a plasma lens located in the intervening galaxy and the Milky Way are shown in Appendix A.

\subsection{Explaining the properties of multiple images in the observations}\label{Sect.3.3}

\begin{figure*}
\centering
  \includegraphics[width=0.99\textwidth, angle=0]{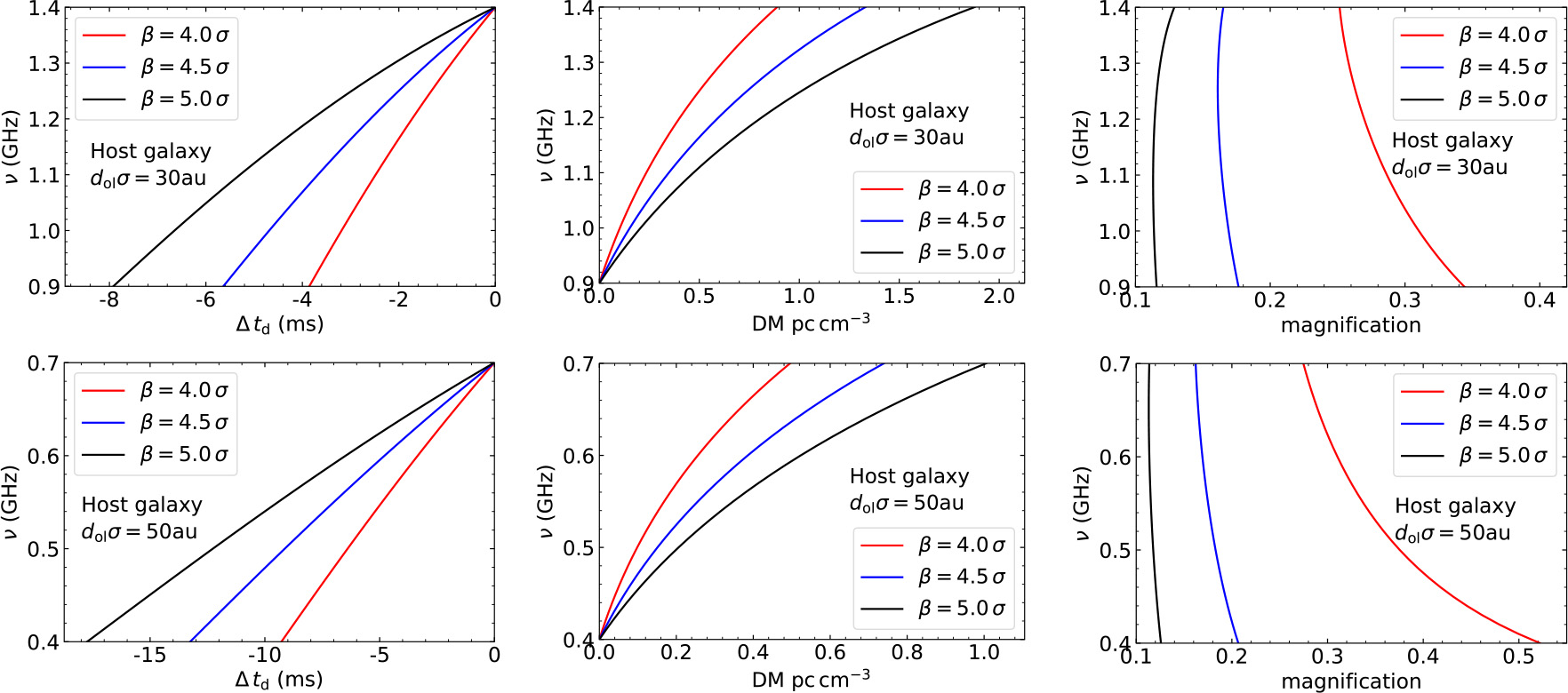}\\
  \caption{Plots showing the delay times (left column), the DMs (middle column) and the magnifications (right column) for the second image at different source positions based on the Gaussian lens in the host galaxy. The different effective structures of the lens are $d_{\rm ol} \sigma = 30 \, \rm au$ (upper panel) and $d_{\rm ol} \sigma = 50 \, \rm au$ (lower panel)}\label{Host}
\end{figure*}

\begin{figure*}
\centering
  \includegraphics[width=0.99\textwidth, angle=0]{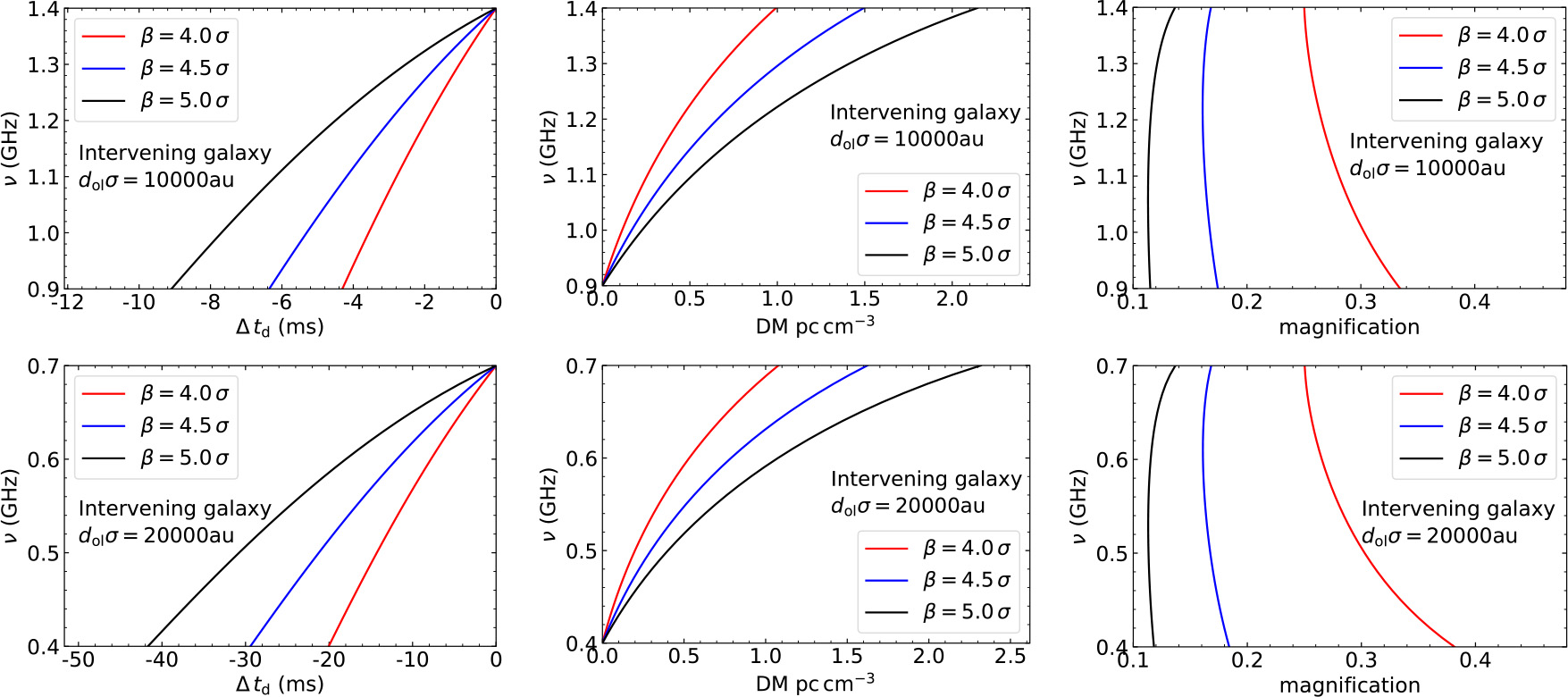}\\
  \caption{Similar to Fig. \ref{Host}, but for a Gaussian lens in an intervening galaxy. The effective structures of the lens are $d_{\rm ol} \sigma = 10^4 \, \rm au$ (upper panel) and $d_{\rm ol} \sigma = 2 \times 10^4 \, \rm au$ (lower panel), respectively}\label{Interval}
\end{figure*}

From the results in the previous subsections, a plasma lens will give rise to different DMs, delay times and magnifications as obtained from the multiple images between the low and high frequencies.
The properties of the second image can be related to observations of FRB 121102 and FRB 180916.J0158+65 (\citealt{Amiri2020,Platts2021}).
Figs. \ref{Host} and \ref{Interval} show the delay time, $\rm DM(\theta)$ and magnification of the second image as due to a Gaussian lens at different distances. From the two figures, the delay time measured at $0.9 \, \rm GHz$ frequency is shorter by several milliseconds and the corresponding DM is lower by $1-2 \, \rm pc \, cm^{-3}$ than that at $1.4 \, \rm GHz$ frequency.
The differences in the delay time and DM between 0.4 and 0.7$\, \rm GHz$ are much higher than $10 \, \rm ms$ and $0.5 \, \rm pc \, cm^{-3}$, respectively.
It is clear from Fig. \ref{Host} and \ref{Interval} that an observer's position closer to the axis of symmetry of the lens will receive radio signals with greater magnifications, but with the value being less than 1.
The differences in delay and DM between high and low frequencies as predicted in our model is consistent with the observations of FRB 121102 and FRB 180916.J0158+65 (\citealt{Amiri2020,Platts2021}).
However, bursts in some repeating FRBs, especially FRB 121102 and FRB 180916.J0158+65, still appear ``sad trombone'', ``happy trombone'' or ``sad trombone'' plus ``happy trombone'' in the frequency-time plot after de-dispersion (\citealt{Amiri2020,Amiri2021,Platts2021}).
A possible reason is that the geometric effects due to the plasma lens were ignored in the de-dispersion, which will be discussed in the next paragraph.

\begin{figure*}
\centering
  \includegraphics[width=0.70\textwidth, angle=0]{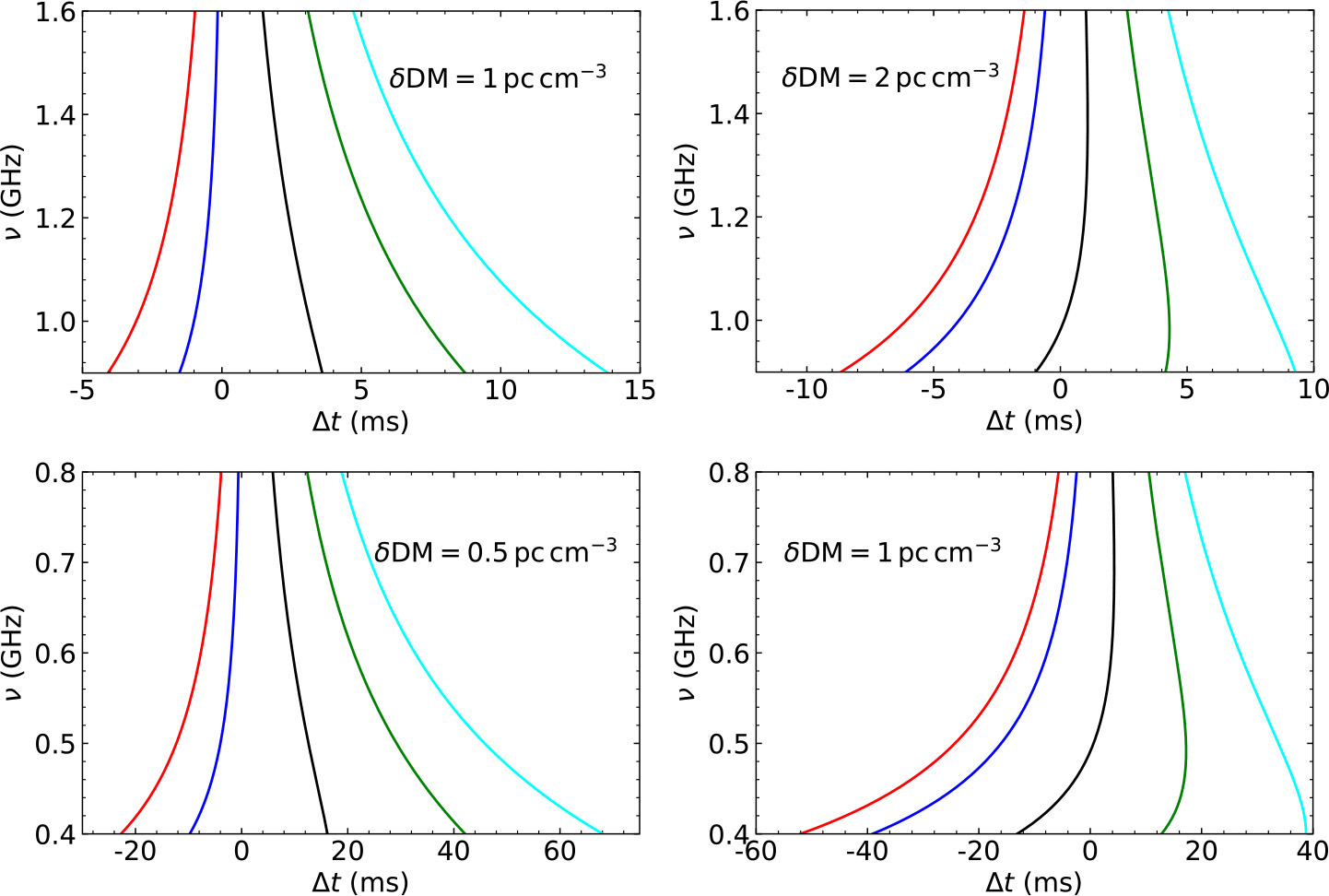}\\
  \caption{Different de-dispersion methods used for radio signal at $0.9-1.6 \, \rm GHz$ (upper panel) and $0.4-0.8 \, \rm GHz$ (lower panel) frequencies. The de-dispersion with $\rm DM = -0.5, \, \, 0, \,\, 1, \,\, 2$, and $3 \rm \, pc \, cm^{-3}$ are indicated by the lines in red, blue, black, green and cyan, respectively.}\label{DM_delay}
\end{figure*}

The DM is estimated by fitting the frequency-time delay curve of the radio sources with the assumption that the density gradient is invariant (\citealt{Petroff2016}).
From Fermat's principle, the inhomogeneous density gradient will contribute to the different propagation paths taken by the background radio signal.
The increase in the delay time as shown in Equations (\ref{geometric_delay}) and (\ref{dispersive_delay}) leads to deviation from the general frequency-time delay relation.
Such effects were discussed by \citet{Er2020,Er2022} in the plasma lens with $\theta_0 < \theta_{0, \rm cr}$, and were also used in pulsars (\citealt{Main2018}) but for lens with $\theta_0 > \theta_{0, \rm cr}$.
Similar to weaker plasma lens with only one image (\citealt{Er2020}), the delay times in the first and third images exhibit the relation of a ``sad trombone'' on the frequency-time plot.
However,  the delay time in the second image shows ``happy trombone'', which can be seen in Figs. \ref{Host} and \ref{Interval}, and the DM is lower at low frequency than that at high frequency.
As mentioned by \citet{Lin2021}, the higher order effects of a perturbed DM with shifting of the line of sight may be required in the theoretical prediction for the delay time. The relationship can be approximated to
\begin{eqnarray}
t(\nu) = 4.15 \, {\rm ms} \, \bigg(\frac{\rm DM}{\nu^{2}_{\rm GHz}}\bigg) - b \, \bigg(\frac{\rm \delta DM^2}{\nu^{4}_{\rm GHz}}\bigg), \label{delay_time1}
\end{eqnarray}
where the first term on the right-hand side of Equation (\ref{delay_time1}) stands for the general frequency-time delay relation, and the different DMs are obtained from different de-dispersion methods.
The $\rm DM = 0$ is the true frequency-time delay relation, whereas $\rm DM > 0$ and $\rm DM < 0$ represent incomplete and excessive de-dispersion signals, respectively.
The second term in Equation (\ref{delay_time1}) represents the geometric effect of a plasma lens due to the perturbed DM, where $b$ is a free parameter, which is assumed $b=1 \, \rm ms$, and $\delta \rm DM$ approximates the difference in perturbed DM between high and low frequencies.
Based on our results and the observations from \citet{Chamma2020} and \citet{Platts2021}, $\delta \rm DM$ at $0.9-1.6 \rm \, GHz$ frequencies can be taken as $1 \rm \, pc \, cm^{-3}$ and $2 \rm \, pc \, cm^{-3}$, and we adopt $0.5 \rm \, pc \, cm^{-3}$ and $1 \rm \, pc \, cm^{-3}$ at $0.4-0.8 \, \rm GHz$ frequencies.
The de-dispersion with DM $ = -0.5, \, \, 0, \,\, 1, \,\, 2$, and $3\rm \, pc \, cm^{-3}$ are used to fit Equation (\ref{delay_time1}). Fig. \ref{DM_delay} shows the frequency-time delay relations.
A radio signal with much smaller de-dispersion DM (than the true DM) forms either ``sad or happy trombone'', whereas $\rm DM \le 0$ gives only ``happy trombone''.
The delay time with some incomplete de-dispersion methods first shows an increase as the frequency decreases, reaching a maximum value, then followed by a decrease.

\begin{figure*}
\centering
  \includegraphics[width=0.6\textwidth, angle=0]{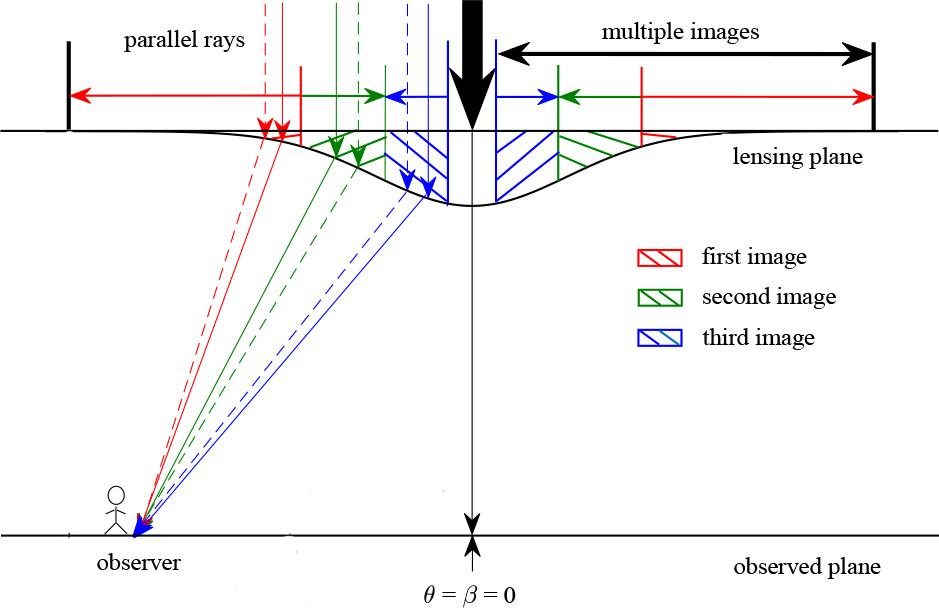}\\
  \caption{Parallel rays passing though a Gaussian lens are diverged to multiple images. The dashed lines stand for high frequency lights detected by the telescope, and the solid lines represent low frequency radio signals.}\label{Lens}
\end{figure*}

It is clear from Section 3.2 that the refracted images are due to signals propagating along different paths of different DMs at the lens plane.
The DMs obtained from the multiple images after the de-dispersion should satisfy $\rm DM_{3}> DM_{2}> DM_{1}$, where the subscripts ``1'', ``2'' and ``3'' stand for the first, second and third images, respectively, in Figs. \ref{delay_x_solution} and \ref{delay_x1_solution}.
Therefore, the distribution of DMs obtained from the signals may contain multiple peaks.
Since the DM from each image is frequency dependent ($\rm DM = DM (\nu)$, e.g., ``sad trombone'' or ``happy trombone''), chromatic deflection occurs at the different bands as illustrated in Fig. \ref{Lens}.
For a repeating FRB, its bursts from the region of the first or third image may have lower DM at high frequency than that at low frequency ($\rm DM_{1,\rm high} < DM_{1, \rm low}$ or $\rm DM_{3,\rm high} < DM_{3,\rm low}$). However, it is the opposite for the radio signals from the region of the second image ($\rm DM_{2,\rm high} > DM_{2,\rm low}$).
Thus the DM differences between two images at two different frequency bands can either be $\Delta \rm DM_{21}(\nu_{\rm high}) > \Delta \rm DM_{21}(\nu_{\rm low})$ or $\Delta \rm DM_{32}(\nu_{\rm high}) < \Delta \rm DM_{3 2}(\nu_{\rm low})$, where $\Delta \rm DM_{21} = DM_2 - DM_1$ and $\Delta \rm DM_{32} = DM_3 - DM_2$.
The time interval of burst pair for some repeating FRBs may be influenced by the properties of the DM differences between images which can be generalized in the form given by
\begin{eqnarray}
\Delta t(\nu) = 4.15 \, {\rm ms} \, \bigg(\frac{\rm \Delta DM}{\nu^{2}_{\rm GHz}}\bigg) \pm b \, \bigg(\frac{\rm \delta DM^2}{\nu^{4}_{\rm GHz}}\bigg), \label{delay_time2}
\end{eqnarray}
where $\Delta \rm DM$ represents the DM differences ($\Delta \rm DM_{21}$, $\Delta \rm DM_{31}$ or $\Delta \rm DM_{32}$), and $\delta \rm DM$ is derived from the perturbation of the geometric effect.
If both $\Delta \rm DM$ and $\delta \rm DM$ are frequency dependent, the similar delay time should result, as shown in Fig. \ref{Lens}.
On the contrary, if they are not frequency dependent, it is straightforward to show that the delay time at GHz frequencies is dependent on the first term on the right-hand side of Equation (\ref{delay_time2}), such that $\Delta t \propto \Delta \rm DM$.
Consider FRB 121102 as an example. The drift rates obtained from different bursts appear to be linearly related to the center frequency of different observing bands ($\partial_{t} \ln(\nu) \propto \nu$) (\citealt{Josephy2019}).
The DM differences between the images are $\Delta \rm DM \approx \chi \nu^2$, with $\chi$ being a constant.
However, the geometric effect remains in Equation (\ref{delay_time2}), and the properties of delay times and DMs as seen in Fig. \ref{DM_delay} are still manifested in the signals.
For a burst pair in FRB 180916.J0158+65, the difference in the arrival time at $400 \, \rm MHz$ frequency is $\sim$$\rm 23 \, \rm ms$ and the drift rate is approximately $-4.2 \, \rm MHz \, ms^{-1}$, and the delay time with the ``sad trombone'' is retained in the subsequent bursts (\citealt{Chawla2020}).
The subsequent bursts may be an incomplete de-dispersion signal, and its delay time for $100\, \rm MHz$ bandwidth is approximately $23.8 \, \rm ms$.
From Equations (\ref{delay_time1}) and (\ref{delay_time2}), the true time interval of the two bursts should be $\Delta t > 46.8 \, \rm ms$.

\subsection{The variations in delay time and DM between multiple images}\label{Sect.3.3}

Observed properties between multiple images are important to probe the true model of the plasma lens.
For a given $\beta$, the perturbation of DM is dependent on specific narrow frequency bands, which leads to the DM being significantly deviated from Equation (\ref{DM}) (\citealt{Cordes2017}).
Due to the relative motion between the source and the lens, or between the observer and the lens, the perturbation of DM may appear to exhibit evolution.
We adopt an effective transverse velocity of the source given by $v_{\bot} = 100 \, \rm km \, s^{-1}$ based on the studies of pulsars and the galaxies (\citealt{Manchester2005,Yang2017}).
Thus the time span for the perturbation of DM can be approximated by (\citealt{Cordes2017})
\begin{eqnarray}
t_{\rm per} \approx \bigg|\frac{\Delta \beta \emph{d}_{\rm ol}}{v_{\bot}}\bigg| \approx 0.0474 \rm \,\, yr \bigg|\frac{\Delta \beta \emph{d}_{ol}}{ \rm au}\bigg| \bigg( \frac{\emph{v}_{\bot}}{100 \, \rm km \, s^{-1}}\bigg)^{-1}, \label{time_span}
\end{eqnarray}
where $\Delta \beta$ is the corresponding maximum change in the source position in the multiple images within the window.

From Section 3.2, the third image occupies relatively lower magnification (than 0.1) at $1.3\, \rm GHz$ (for $d_{\rm ol}\sigma = 30 \, \rm au$ or $ 10^4 \, \rm au$) and at $0.6\, \rm GHz$ (for $d_{\rm ol}\sigma = 50 \, \rm au$ or $2\times 10^4 \, \rm au$), which will be ignored.
Fig. \ref{beta_theta_sigma_x} shows that the first and second images possess the same image position at the inner boundary, which implies that the differences in the delay time and the DM are both zero.
As the source position increase gradually, the delay time and DM difference for the two images are also increasing, and their maximal differences are at the outer boundary.
Thus the rate of change in DM difference with time can be approximated by (\citealt{Yang2017})
\begin{eqnarray}
\nonumber \bigg|\frac{d\rm DM}{dt}\bigg| &\approx& |\Delta \rm DM| \bigg|\frac{\Delta \beta \emph{d}_{ol}}{\emph{v}_{\bot}}\bigg|^{-1} \approx 21.081 \rm \,pc \, cm^{-3} \, yr^{-1}\\
&\times& \bigg|\frac{\Delta \rm DM}{ \rm  \, pc\, cm^{-3}}\bigg| \bigg|\frac{\Delta \beta \emph{d}_{\rm ol}}{\rm au}\bigg|^{-1} \bigg( \frac{\emph{v}_{\bot}}{100 \, \rm km \, s^{-1}}\bigg),\label{dDM}
\end{eqnarray}
where $\Delta \rm DM$ is the maximum variation in the DM difference.
The corresponding rate of change for the delay time difference approaches to
\begin{eqnarray}
\bigg|\frac{\Delta t}{t_{\rm per}}\bigg| \approx 21.097 \, {\rm ms \, yr^{-1}} \, \bigg|\frac{\Delta t}{\rm ms}\bigg| \bigg|\frac{\Delta \beta \emph{d}_{\rm ol}}{\rm au}\bigg|^{-1} \bigg( \frac{\emph{v}_{\bot}}{100 \, \rm km \, s^{-1}}\bigg){\bg ,} \label{dtime}
\end{eqnarray}
where $\Delta t$ is the maximum variation in the delay time difference.

Table \ref{tb1} gives the variations in the DM difference and the delay time difference for burst pairs and their time spans due to the effects of a Gaussian lens. It is apparent from the table that the rates of change for the differences in the delay time and in the DM should be relatively large when the plasma lens is located in the host galaxy, or in the Milky Way, with the effects lasting for about several years.
Conversely, the effects of a plasma lens in intervening galaxy are at a much longer timescale of $1000 \, \rm yr$, but their effects are not significant over several years.

\begin{table*}
  \centering
  \caption{Variations in the values of specific parameters between the first and second images for the Gaussian lens at $0.6 \rm \, GHz$ and $1.3 \rm \, GHz$ frequencies. The plasma lenses in the host galaxy and in the Milky Way both occupy a structure scale of $30 \, \rm au$ or $50 \, \rm au$, whereas it is $10000 \, \rm au$ or $20000 \, \rm au$ for the plasma lens in the intervening galaxy.}\label{tb1}
  \begin{center}
  \begin{tabular}{lcccccc}
  \hline\hline\noalign{\smallskip}
  Structure Scale                                      & $30 \, \rm au$ (Host) & $30 \, \rm au$ (MW) &$10000 \, \rm au$  & $50 \, \rm au$ (Host) & $50 \, \rm au$ (MW) & $20000 \, \rm au$  \\
  Referred Frequency                                   &$1.3 \, \rm GHz$& $1.3 \, \rm GHz$   & $1.3 \, \rm GHz$   &$0.6 \, \rm GHz$ &$0.6 \, \rm GHz$ & $0.6 \, \rm GHz$   \\
  \hline\noalign{\smallskip}
  $|\Delta \beta| \, \, (\sigma)$                      & $2.081$        & $1.862$   & $2.287$            & $1.569$     & $1.542$   & $1.653$            \\
  $|\Delta \rm DM| \, \, (\rm pc \, cm^{-3})$          & $2.247$        & $1.500$   & $2.784$            & $0.371$     & $0.294$   & $0.664$            \\
  $\Delta t \, \, \rm (ms)$                            & $22.926$       & $15.457$  & $28.513$           & $33.896$    & $27.067$  & $59.162$            \\
  $t_{\rm per} \, \, (\rm yr)$                         & $2.959$        & $2.648$   & $1083.868$         & $3.718$     & $3.655$   & $1567.082$         \\
  $|d{\rm DM}/dt| \, \, (\rm pc \, cm^{-3} \, yr^{-1})$& $0.75874$      & $0.56593$ & $0.00257$          & $0.09977$   & $0.08031$ & $ 0.00042$          \\
  $|d(\Delta t)/dt|  \, \, (\rm ms \, yr^{-1})$        & $7.749$        & $5.837$   & $0.026$            & $9.117$     & $7.408$   & $0.038 $           \\
  \noalign{\smallskip}\hline\hline
  \end{tabular}
  \end{center}
\end{table*}

\section{Discussion and Summary}\label{Sect.5}

We have shown that the properties of a plasma lens and their relative distances to an observer play a leading role in the formation of multiple images from FRBs.
The separated images with significant frequency-dependent time delays are caused by a plasma lens.
We demonstrate that the delay times of the first and third images are shorter at high frequency than these at low frequency.
However, the radio signals of the second image should arrive at the telescope earlier at low frequency than at high frequency.
This is due to the the geometric effects of plasma lens, which gives rise to the inversed frequency-time delay relation in the second image.
The variation of DM, the time intervals between the images and their time spans due to the motion of the source relative to the plasma lens are significant in the host galaxy but less so in the intervening galaxy.

In this study, the thin lens approximation has been adopted.
Due to the magnification limit ($\ge0.1$), the radio signal in relation to the multiple images seems to show narrow-band spectrum, and a radio telescope may detect the two images as burst pair.
The time interval between multiple images within the detection window should depend on the source position and the narrow-band spectrum, which is consistent with some burst pairs from the repeaters (\citealt{Chawla2020,Platts2021}).
However, the predicted time interval between multiple images using a single or simple lens plane is not sufficient to account for the large range of interval times for burst pairs.
In addition, the repeating bursts occur at relatively short burst rate (\citealt{Andersen2019,Fonseca2020}) and the magnifications of some images are independent of the effect of the plasma lens.
The waiting time between two adjacent bursts in a continuous observation may emerge as separated multiple distributions and irrelevant to the high energy components of bursts because of the effect of the plasma lens (\citealt{Li2021b}).

The FRB dispersion relation can be influenced by the inhomogeneous properties of the plasma along the line of sight, which leads to deviation from the classical dispersion relation.
This chromatic effect due to plasma lenses may exist at all distance scales and is a very important tool to reveal the dispersion relation.
Based on the large sample of FRBs at $600 \, \rm MHz$, an event rate of $818 \, \rm sky^{-1} \, day^{-1}$ has been inferred above a fluence of $5 \, \rm Jy \, ms$ (\citealt{Amiri2021}).
The DM as derived from pulsars based on the interstellar medium in the Milky Way ranges from 3 to $1700 \, \rm pc \, cm^{-3}$, with the largest DM expected around the galactic disk (\citealt{Manchester2005}).
FRBs may traverse foreground objects similar to the Milk Way before reaching the observer (\citealt{Fedorova2019,Prochaska2019,Xu2021}). This implies that a repeater, in particular the ones that located at different regions of an intervening galaxy, may form multiple DM distributions.
Thus the detailed dispersion properties of FRBs can be used to research the properties of near-source plasma and the intervening galaxy, such as the properties of supernova remnants, pulsar wind nebulae, H II regions, black holes surrounded by plasma and galactic halo (\citealt{Yang2017,Prochaska2019,Tsupko2019}).

\begin{acknowledgements}
We would like to thank the XAO pulsar group for discussions and the anonymous referee for helpful suggestions that led to significant improvements in our study.
We are thankful to Prof. XinZhong Er and Prof. Adam Rogers for some useful advice.
The work is supported by the National Natural Science Foundation of China (Grant No.12041304, 11873080, 12033001).
Z.G.W. is supported by the 2021 project Xinjiang Uygur Autonomous Region of China for Tianshan elites, and the National SKA Program of China (Grant No. U1838109, 2020SKA0120100, 12041301).
RY is supported by the Key Laboratory of Xinjiang Uygur Autonomous Region No. 2020D04049, the National SKA Program of China No. 2020SKA0120200, and the 2018 Project of Xinjiang Uygur Autonomous Region of China for Flexibly Fetching in Upscale Talents.
X.Z is supported by CAS ``Light of West China'' Program No. 2018-XBQNXZ-B-025.
The Nanshan 26-m Radio Telescope is partly supported by the Operation, Maintenance and Upgrading Fund for Astronomical Telescopes and Facility Instruments, budgeted from the Ministry of Finance of China (MOF) and administrated by the CAS.

\end{acknowledgements}

\begin{appendix}

\section{The effects of the plasma lens in the intervening galaxy and Milky Way}

\begin{figure*}
\centering
  \includegraphics[width=0.99\textwidth, angle=0]{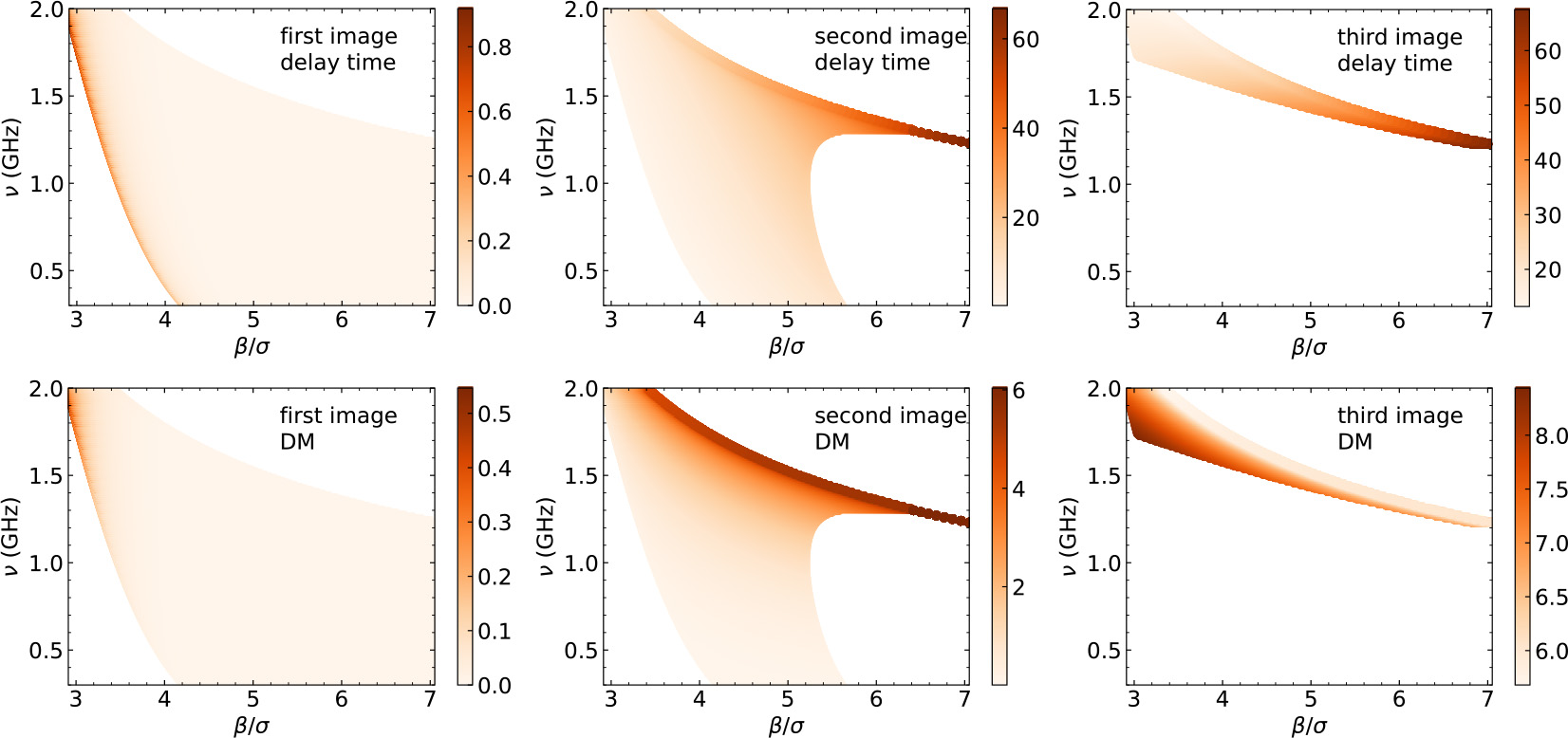}\\
  \caption{Similar to Fig. \ref{delay_x_solution}, the delay time (upper panel) and DM (lower panel) for the three images as caused by the Gaussian lens at $ z_{\rm d} = 0.0219367$ in the intervening galaxy and $d_{\rm ol} \sigma = 10^4 \, \rm au$.}\label{delay_x_solution1}
\end{figure*}

\begin{figure*}
\centering
  \includegraphics[width=0.99\textwidth, angle=0]{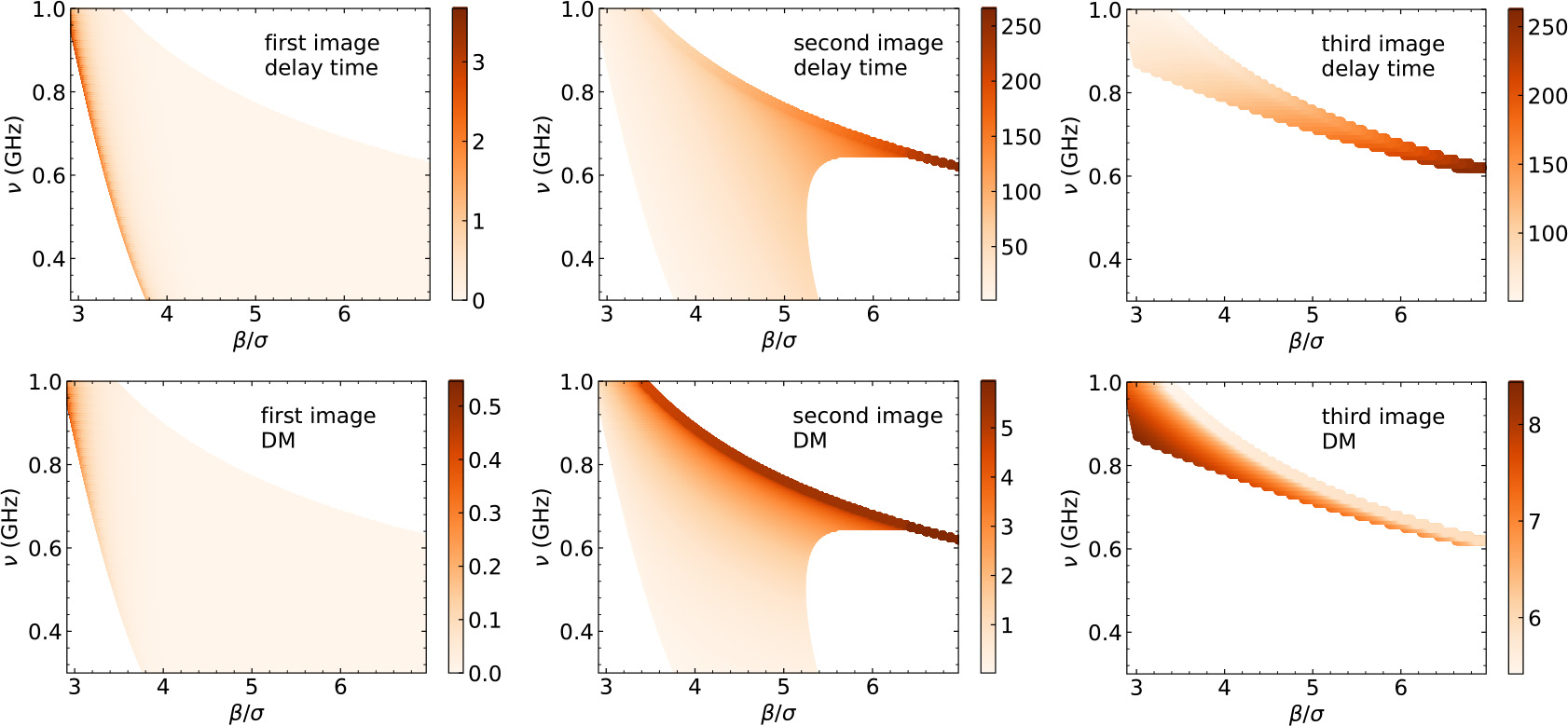}\\
  \caption{Similar to Fig. \ref{delay_x1_solution} and Fig. \ref{delay_x_solution1}, the delay time (upper panel) and DM (lower panel) of the three images as caused by the Gaussian lens in the intervening galaxy ($ z_{\rm d} = 0.0219367$) and with $d_{\rm ol} \sigma = 2 \times 10^4 \, \rm au$.}\label{delay_x1_solution1}
\end{figure*}

\begin{figure*}
\centering
  \includegraphics[width=0.99\textwidth, angle=0]{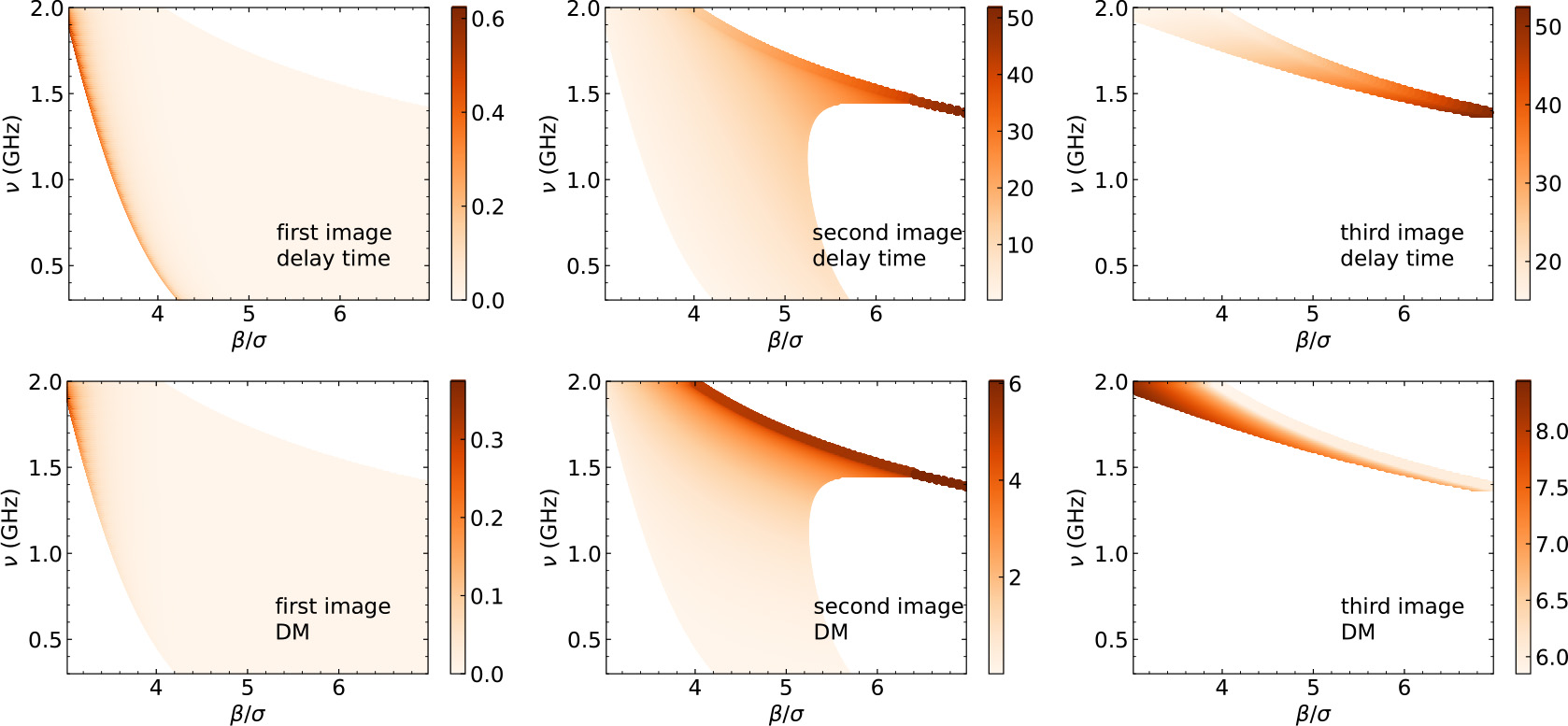}\\
  \caption{Similar to Fig. \ref{delay_x_solution} but for the Gaussian lens located in the Milky Way at $d_{\rm ol} \sigma = 30 \, \rm au$.}\label{delay_x_solution2}
\end{figure*}

\begin{figure*}
\centering
  \includegraphics[width=0.99\textwidth, angle=0]{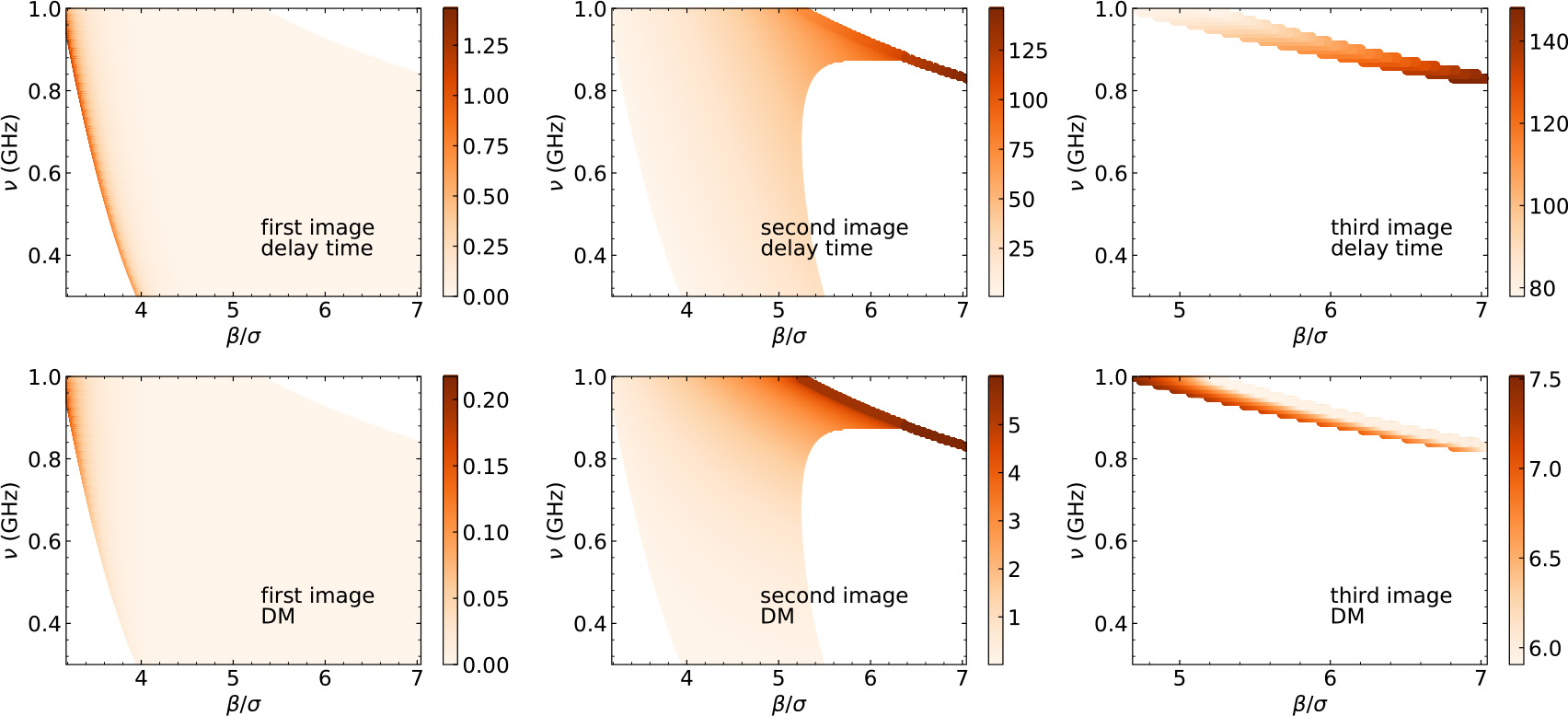}\\
  \caption{Similar to Fig. \ref{delay_x1_solution} but the location of the Gaussian lens is in the Milky Way is at $d_{\rm ol} \sigma = 50 \, \rm au$.}\label{delay_x1_solution2}
\end{figure*}

This appendix gives the DM and delay time for each of the three images due to a plasma lens locating in the intervening galaxy, as shown in Figs. \ref{delay_x_solution1} and \ref{delay_x1_solution1}, and the Milky Way, as shown in Figs. \ref{delay_x_solution2} and \ref{delay_x1_solution2}.
The three images exhibit similar properties as in the host galaxy discussed in Section 3.2, but with several differences in the values of the DM and delay time.

\end{appendix}

\label{lastpage}

\clearpage

\end{document}